\documentclass[prl,amsmath,amssymb, aps,onecolumn,10pt,superscriptaddress]{revtex4-2}
\usepackage{graphicx,xcolor,array,float,amsmath,amsfonts}

\begin{document} 

\title{Continuous-wave, high-resolution, ultra-broadband mid-infrared nonlinear spectroscopy with tunable plasmonic nanocavities} 

\author{Zhiyuan Xie} 
%\affiliation{Institute of Physics, Ecole Polytechnique F\'ed\'erale de Lausanne (EPFL), CH-1015 Lausanne, Switzerland}
\author{Nobuaki Oyamada} 
%\affiliation{Institute of Physics, Ecole Polytechnique F\'ed\'erale de Lausanne (EPFL), CH-1015 Lausanne, Switzerland}
\author{Francesco Ciccarello} 
\author{Valeria Vento}
\affiliation{Institute of Physics, Ecole Polytechnique F\'ed\'erale de Lausanne (EPFL), CH-1015 Lausanne, Switzerland}
\author{Wen Chen}
\affiliation{State Key Laboratory of Precision Spectroscopy, East China Normal University}
\author{Christophe Galland}
\affiliation{Institute of Physics, Ecole Polytechnique F\'ed\'erale de Lausanne (EPFL), CH-1015 Lausanne, Switzerland}
\email{chris.galland@epfl.ch}

\begin{abstract}
\textbf{
Vibrational sum- and difference-frequency generation (SFG and DFG) spectroscopy probes the nonlinear response of interfaces at mid-infrared (MIR) wavelengths while detecting upconverted signals in the visible. Recent work has moved from large-area films and colloids to nanoscale structures using dual-resonant plasmonic nanocavities that co-confine light and matter in deep-subwavelength volumes. 
Here we implement high-resolution ($<1$~cm$^{-1}$), continuous-wave ultrabroadband vSFG, vDFG, and four-wave mixing (FWM) coherent spectroscopy from 860 to 1670~cm$^{-1}$ on dual-resonant antennas under ambient conditions. Using a commercial, broadly tunable quantum-cascade laser and eliminating geometric phase matching simplify acquisition and expand spectral reach.  
The resulting spectra exhibit coherent interference between resonant (vibrational) and nonresonant (electronic) contributions to the effective $\chi^{(2)}$, previously accessible only under fs/ps excitation. Simultaneous measurement of SFG and DFG enables a {ratiometric} analysis that suppresses common-mode drifts and helps reveal vibrational resonances. 
We demonstrate versatility and reproducibility across several analytes that span distinct relative strengths of vibrational vs. electronic nonlinearities. 
Together, these capabilities position our approach as a scalable route to multiplexed, high-resolution MIR sensing and a practical basis for chip-level, label-free coherent spectroscopy. It opens a feasible path toward single- and few-molecule optomechanical studies using nanoscale trapping strategies.
}
\end{abstract}

\maketitle
%%%%%%%%%%%%%%%%%%%%%%%%%%%%%%%%%%%%%%%%%%%%%%%%%%%%%%%%%%%%%%%%I reused all old references and added a bit more. hopefully they are good like this now. in the bib. the last few was newly added. 
\section*{Introduction}
Vibrational absorption spectroscopy in the mid-infrared (MIR, 400–4,000~cm$^{-1}$) and vibrational Raman spectroscopy in the visible (VIS) wavelength domain underpin label-free molecular identification across chemistry, materials, and biology.% by reading out bond-specific fingerprints. 
Field-enhanced implementations, namely, surface-enhanced Raman scattering (SERS) and surface-enhanced infrared absorption (SEIRA), have expanded their sensitivity down to molecular monolayers and nanomolar concentrations in liquids \cite{stuart2004,anker2008,neubrech2017,prater2020}.
Beyond these single-frequency excitation schemes, simultaneous driving of molecular vibrations with MIR and VIS fields coherently interrogates modes that are both IR- and Raman-active through nonlinear frequency mixing.
In vibrational sum- and difference-frequency generation (SFG/DFG), a resonant $\chi^{(2)}$ response arises when the MIR frequency matches a vibrational resonance. Since a centrosymmetric bulk has a vanishing $\chi^{(2)}$, these readouts are intrinsically surface- and interface-specific, and provide rich information on molecular orientation, ordering, etc. \cite{lambert2005,shen2016,pickering2022,shen1989,buck2001,geiger2009,takahashi2025}. Over the past two decades, plasmonic nanoparticles and antenna arrays have provided fertile testbeds for such probes \cite{kawai2000,baldelli2000,humbert2005,weeraman2006,tourillon2007,bordenyuk2007,pluchery2009,li2009,tourillon2009,humbert2011,humbert2013,dalstein2015,humbert2019,busson2019,dalstein2019,linke2019,tan2022,ma2023,pei2024,zheng2024,barbillon2025,pei2025,lis2013,barbillon2018,gao2020,guo2020}, revealing nanoparticle-induced adsorbate reconfiguration \cite{weeraman2006,bordenyuk2007,barbillon2025}, enhanced chiral responses \cite{ma2023}, and altered vibrational relaxation and dephasing \cite{zheng2024,zheng2025,pei2025,bell2025}. More recently, near-field SFG spectroscopy was achieved using tip-enhanced fields \cite{sakurai2025,roelli2025, takahashi2025}, while nonlinear MIR spectroscopy was extended to probe the third-order ($\chi^{(3)}$) response through 2-MIR photon four-wave mixing under picosecond pulsed excitation \cite{liu2025,bell2025}.

In coherent vibrational spectroscopy, the experimentally accessible spectral window is often limited by the available ultrafast laser wavelengths and the need to maintain geometric phase-matching. Most broadband implementations rely on femtosecond or picosecond sources and macroscopic ensembles, with modest plasmonic enhancements and strict phase-matching conditions. 
As a result, an integrated approach that would allow routine coherent nonlinear spectroscopy within arbitrary MIR sub-bands, with sub-micron spatial resolution, and simultaneous access to the Raman spectrum has remained challenging.

%%%%%%%%%%%%%%%%%%%%%%%%%%%%%%%%%%%%%%%%%%%%%%%%
\begin{figure}[h]
        \centering
        \includegraphics[width=1\linewidth]{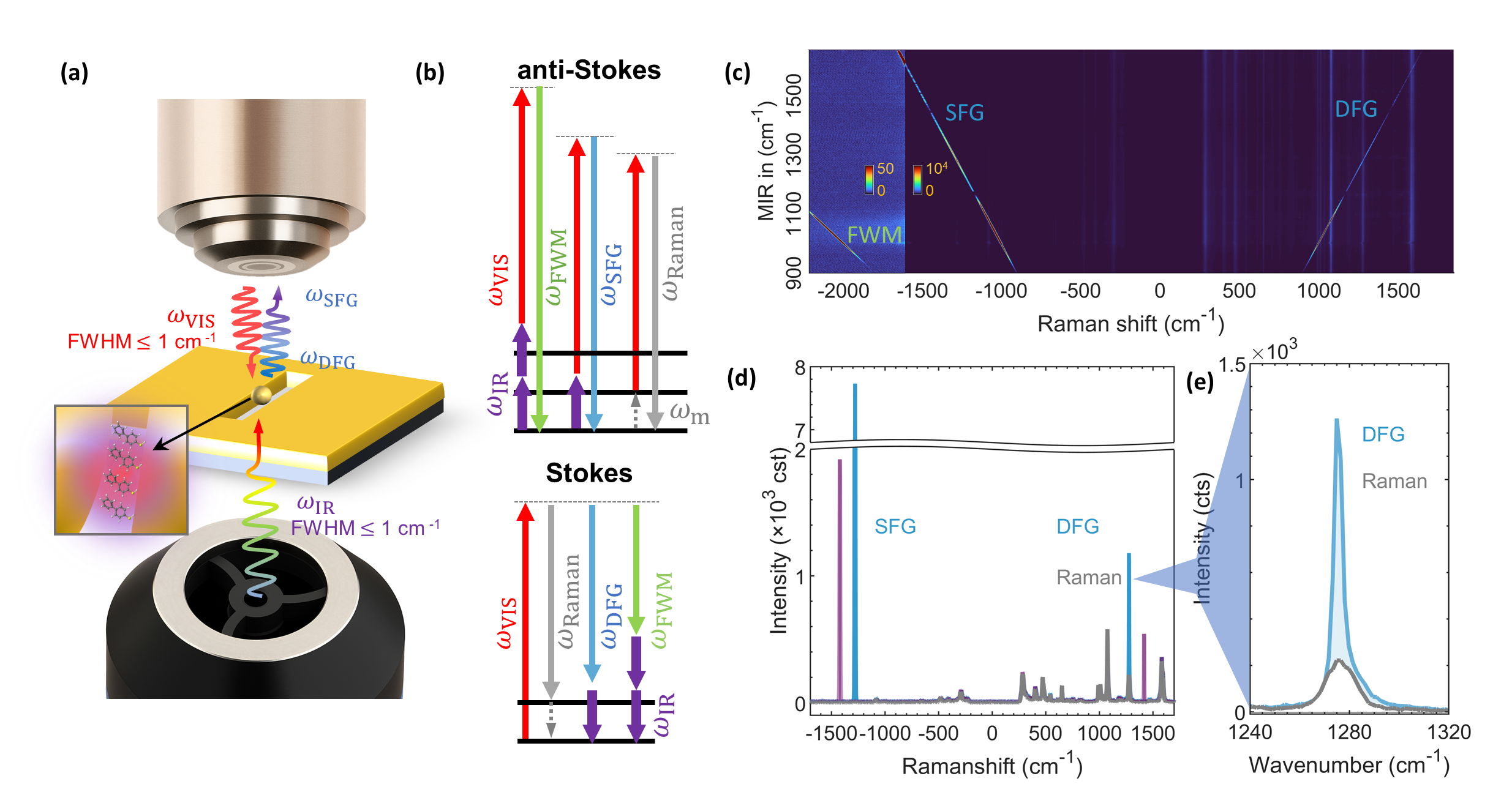}
     \caption{
\textbf{(a)} Experimental scheme. A 780\,nm VIS laser (objective NA = 0.9) and a tunable MIR QCL (NA = 0.78) are co-focused onto a nanoparticle-on-slit (NPoS) nanocavity. The analyte layer resides in the nanometer gap between the gold nanoparticle and the slit walls (inset). \textbf{(b)} Energy-diagram representation of photon scattering pathways under simultaneous VIS \((\omega_{\mathrm{VIS}})\) and MIR \((\omega_{\mathrm{IR}})\) excitation. 
\textbf{(c)} Two-dimensional map of cavity emission during a continuous MIR sweep (890--1650\,cm\(^{-1}\), 1\,cm\(^{-1}\) steps; 761 frames; vertical axis) from an individual BPhT-functionalised NPoS (from M1 array, see below). The SFG, DFG, and 2p-FWM peaks are labeled.   
\textbf{(d)} Representative emission spectra at selected MIR wavelengths (coloured), compared with the MIR-off Raman spectrum (grey), all from the same nanocavity. \textbf{(e)} Zoom-in on the Stokes-sideband region around the DFG signal. After subtracting spontaneous Raman (grey), the net DFG peak is fit with a Gaussian, whose area is $\tilde{I}^{(2)}_{\pm}(\omega_{\mathrm{IR}})$ in eq.~(\ref{eq:normI}).
}
        \label{fig:concept}
    \end{figure}

%%%%%%%%%%%%%%%%%%%%%%%%%%%%%%%%%%%%%%%%%%%%%%%%%%%%%%%%%%%%%%%%

Here, building on recent developments in dual-resonant nanocavity designs \cite{chen2021,hu2025,xomalis2021,bell2025},  we introduce a versatile, continuous-wave (cw) platform that removes the experimental constraints on tunable ultrafast laser sources and geometric phase-matching, detailed in \cite{Xie2026}. 
A 780~nm VIS laser (C-WAVE GTR), a tunable MIR quantum cascade laser (MIRcat QCL), and an addressable-slit metasurface are combined so that focusing the MIR and VIS beams onto a single slit co-localizes their near fields in a doubly resonant nanoparticle-on-slit cavity \cite{chen2021,hu2025}. This colocalized, sub-wavelength interaction eliminates the need for geometric phase-matching and drives efficient $\chi^{(2)}$ upconversion at low cw powers. From the same nanocavity, we simultaneously retrieve four signal channels, namely SFG, DFG, two-MIR photon four-wave mixing (2p-FWM), and SERS, each probing distinct and complementary aspects of the interfacial optical response. The operational MIR band can be adapted by selecting the QCL gain chips and the gold nanoslit length and periodicity, which enables ultrabroadband, continuous scans from 860 to 1670~cm$^{-1}$ with sub-wavenumber, laser-defined resolution, under ambient conditions. Simultaneous SFG and DFG acquisition enables a ratiometric readout that suppresses common-mode drifts and isolates vibrational resonances in the effective $\chi^{(2)}$ response. The signals are sufficiently strong that, in addition to spectrometer readout, a single-pixel detector behind a band-pass filter can retrieve the vSFG spectrum, which points to simplified instrumentation options.

%%%%%%%%%%%%%%%%%%%%%%%%%%%%%%%%%%%%%%%%%%%%%%%%%%%%%%%%%%%%%%%%
\section*{Results}

\paragraph{Principle of the experiment}
A doubly resonant nanoparticle-on-slit (NPoS) cavity \cite{chen2021,hu2025} concentrates MIR and VIS fields into the nanometre gap between a gold nanoparticle (150\,nm nominal diameter) and the slit walls, which boosts nonlinear frequency mixing from only a few molecules (Fig.~\ref{fig:concept}a,b). We pattern dense arrays of slits that function as MIR metasurfaces (process in Fig.~S1). The slit {length} sets the single-antenna MIR resonance \cite{hu2025}, whereas the {array periodicity} shapes the collective MIR absorption envelope measured by FTIR (Fig.~S2). Thus, by choosing the length and pitch, we arbitrarily design the MIR band of the sensor, independent of the VIS excitation wavelength.
A tunable MIR QCL is focused through the Si substrate by a reflective Cassegrain objective (NA 0.78, Pike Technologies; Fig.~S3), together with a 780\,nm VIS beam focused from the other side with a high-NA objective (NA 0.9) also used to collect all signals shown below. Under simultaneous illumination of a single nanocavity, the inelastic signal contains the four nonlinear processes described in Fig.~\ref{fig:concept}b (SFG, DFG, 2p-FWM and SERS).

A representative full two-dimensional dataset is shown in Fig.~\ref{fig:concept}c. On the horizontal axis is the detection frequency with respect to the VIS laser line; we adopt the Raman convention in which negative shifts denote anti-Stokes emission at shorter wavelengths.
The vertical axis corresponds to the MIR frequency being stepwise swept.
The two-dimensional map reveals sharp peaks following different slopes in the Raman-shift–versus–$\omega_{\mathrm{IR}}$ plane, consistent with DFG, SFG (slope of $\pm 1$) and 2p-FWM (slope of $\pm -2$).
Our spectrometer is configured so that both $\chi^{(2)}$ sidebands are recorded simultaneously: SFG appears on the anti-Stokes side of the VIS laser line, and DFG appears on the Stokes side, as seen in Fig.~\ref{fig:concept}c.  Because of the limited spectrometer range, 2p-FWM is recorded on the anti-Stokes side only, but we can also measure it on the Stokes side.  
Fig.~\ref{fig:concept}d shows two single-shot (1\,s exposure) emission spectra acquired at two different MIR pump wavelengths %(blue line by xcm$^{-1}$, purple line for xcm$^{-1}$, and grey line for no MIR pump) 
under identical VIS excitation conditions. In some scans, the SFG/DFG coverage is incomplete. This effect is not intrinsic to the metasurface or to the analyte. It arises when the MIR power drops due to a swap of gain chip, or due to a narrow atmospheric water absorption line (see IR power spectrum in Fig.~S4).

%%%%%%%%%%%%%%%%%%%%%%%%%%%%%%%%%%%%%%%%%%%%%%%%%%%%%%%%%%%%%%%%

    \begin{figure}[h] % the position of the figure, does it matter?
        \centering
        \includegraphics[width=1\linewidth]{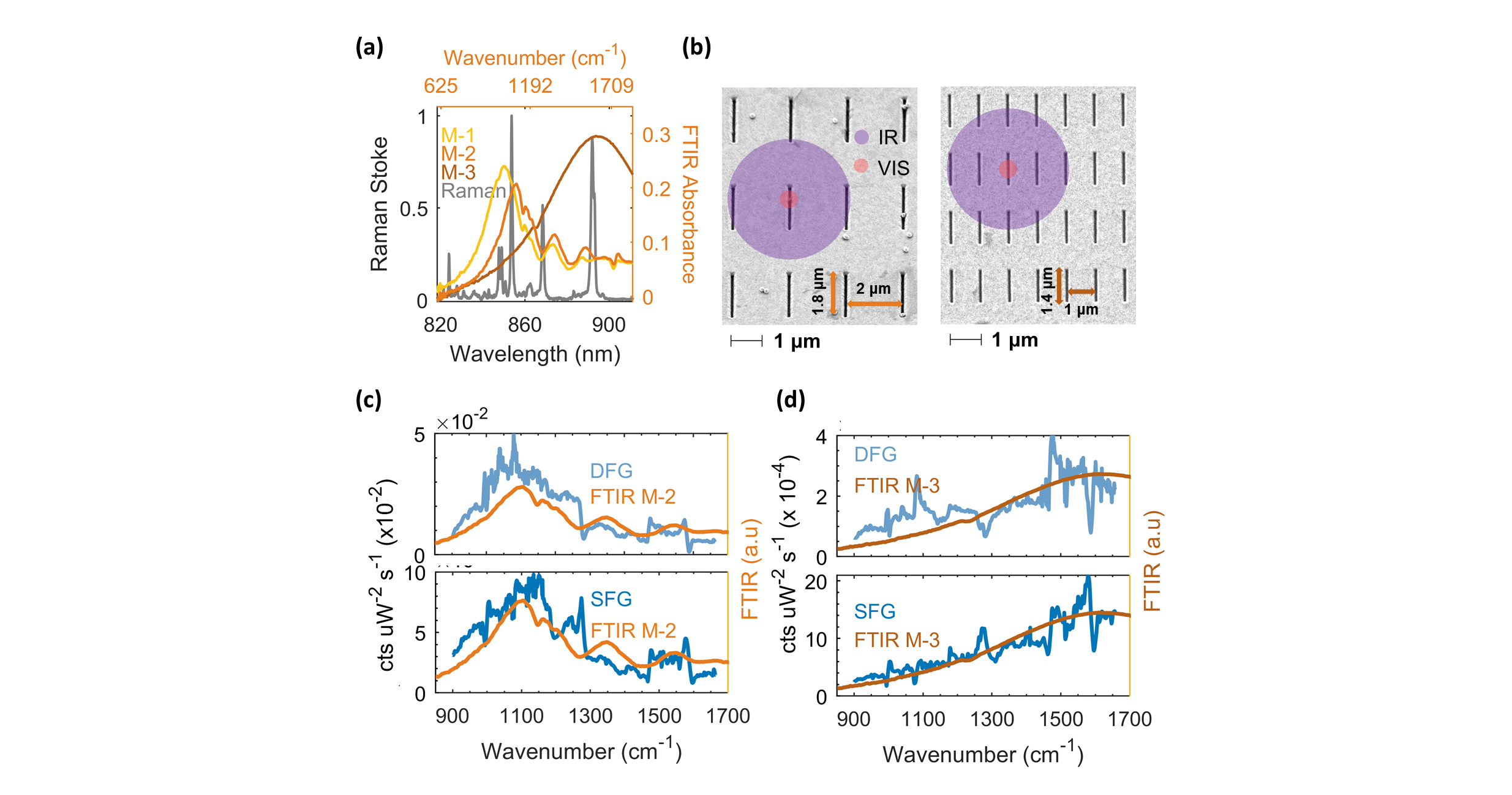}
       \caption{\textbf{(a)} Fourier-transform IR (FTIR) absorption spectra of three metasurface designs (M-1/M-2/M-3, right axis), overlaid with the 780~nm Raman spectrum of BPhT (left axis).\textbf{(b)} Scanning-electron microscope (SEM) images of two representative slit-array MIR metasurfaces: M-2 (left) and M-3 (right). An estimate of the VIS and MIR illumination spot sizes are overlaid on each image. \textbf{(c–d)} Power-normalized DFG (top, light blue) and SFG (bottom, dark blue) spectra from single BPhT-functionalized NPoS cavities fabricated on the M-2 (c) and M-3 (d) metasurfaces. For both designs, the nonlinear signal strength follows the MIR absorption spectrum of the corresponding array (orange curves). 
       }
        \label{fig:arrays}
    \end{figure}
%%%%%%%%%%%%%%%%%%%%%%%%%%%%%%%%%%%%%%%%%%%%%%%%%%%%%%%%%%%%%%%%.

 For each MIR wavenumber $\omega_{\mathrm{IR}}$, we record the MIR power $P_{\mathrm{IR}}(\omega_{\mathrm{IR}})$ just before the sample. Frames with $P_{\mathrm{IR}}$ below a fixed detection threshold are excluded from the data. 
To analyse the nonlinear response, we extract the net SFG and DFG signal intensities at each MIR wavenumber, $\tilde{I}^{(2)}(\pm \omega_{\mathrm{IR}})$, by subtracting the spontaneous Raman background from the total emission (Fig.~\ref{fig:concept}d; see also section S4 of the SI). Below, we always report the normalized quantity  %%% remember to add this figure into SI
\begin{equation}\label{eq:normI}
{I}^{(2)}(\pm \omega_{\mathrm{IR}})=\frac{\tilde{I}^{(2)}(\pm \omega_{\mathrm{IR}})}{P_{\mathrm{VIS}}\,P_{\mathrm{IR}}(\omega_{\mathrm{IR}})}\!,
\end{equation}
where $P_{\mathrm{VIS}}$ is the VIS power on sample. The MIR wavenumber step is typically set as 1\,cm$^{-1}$; with an accuracy down to 0.5\,cm$^{-1}$ per manufacturer data.
Because third-order nonlinear response is not restricted by inversion symmetry, the metal bulk (within the field penetration depth) can contribute to the 2p-FWM \cite{liu2025} while SFG and DFG are interface-specific. 
We sometimes observe an increase in the anti-Stokes Raman background when the 2p-FWM signal is stronger. This behaviour is consistent with increased MIR absorption that drives both SFG/DFG and FWM processes, resulting in moderate substrate heating ($<30$~K).

\paragraph{Broadband, high-resolution SFG and DFG from single NPoS nanocavities}
Figure~\ref{fig:arrays}a overlays the FTIR absorption of three metasurfaces with the 780~nm SERS spectrum of biphenyl-thiol (BPhT) recorded from a single NPoS. The SERS spectrum is highly reproducible and remains invariant across individual cavities from different metasurfaces, indicating comparable molecular coverage and a stable VIS plasmonic response. 
Figures~\ref{fig:arrays}c–d show two representative datasets in which power-normalized SFG and DFG spectra are retrieved from two individual BPhT-functionalized NPoS placed on metasurfaces M-2 and M-3, respectively, each characterized by distinct MIR absorption spectra. 
There is an obvious correspondence between the slowly varrying envelope of the nonlinear signal strength and the IR absorption spectrum of the supporting metasurfaces. This observation is consistent with the fact that DFG and SFG originate from the resonantly enhanced nanocavity near-field \cite{hu2025}. Vibrational modes of the BPhT molecules give rise to resonances in the molecular $\chi^{(2)}$, which result in sharp dispersive perturbations in the detected SFG and DFG. This is due to the interference of this resonant signal with an approximately constant nonresonant background, $\chi^{(2)}_{\rm nr}$, resulting from far detuned electronic transitions.
At the peak of the MIR resonance, the power-normalized single-cavity SFG signal intensity typically reaches $10^{5}$~cts/s/mW$^2$ for M-2 and $10^{4}$~cts/s/mW$^2$ for M-3. 
Such cavity-resolved, broadband nonlinear response was inaccessible in earlier implementations that lacked wide MIR tunability \cite{chen2021} or were limited by incoherent readout and insufficient spectral resolution \cite{xomalis2021}.

%%%%%%%%%%%%%%%%%%%%%%%%%%%%%%%%%%%%%%%%%%%%%%%%%%%%%%%%%%%%%%%%

    \begin{figure}[h] % the position of the figure, does it matter?
        \centering
        \includegraphics[width=0.95\linewidth]{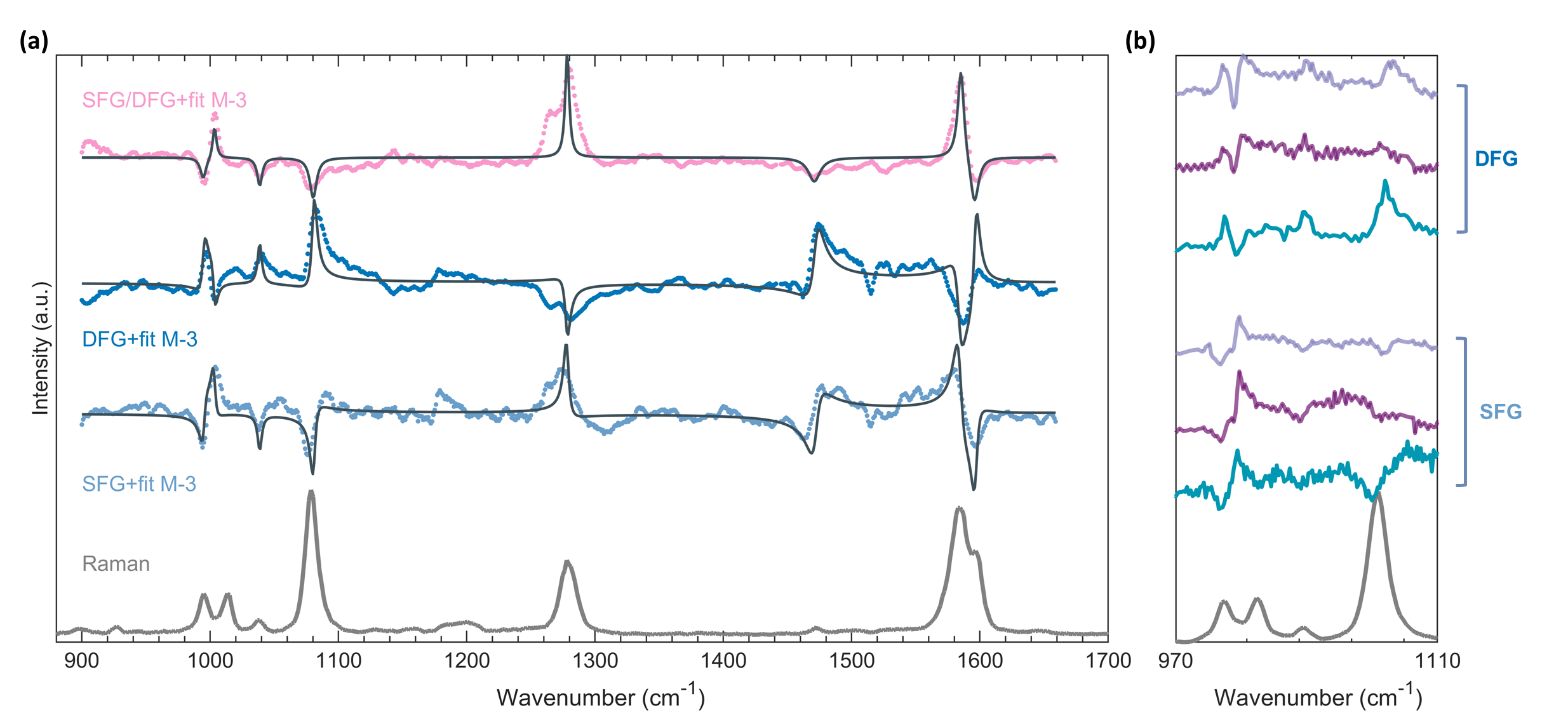}
        \caption{\textbf{(a)} Single–nanocavity broadband readout. From bottom to top: a representative Stokes Raman spectrum (grey) used as a vibrational reference (invariant across NPoS); power-normalized SFG and DFG spectra (blue dots) from the NPoS in Fig.~\ref{fig:arrays}d together with their fits (black lines); and the corresponding SFG/DFG ratio (pink dots) from the same cavity. Fit parameters are summarized in Table~S2. \textbf{(b)} SFG and DFG spectra (colore coded) from three independent NPoS on two metasurfaces (M-2, M-3), overlaid with a typical SERS spectrum (grey).}
        \label{fig:broadband}
    \end{figure}
%%%%%%%%%%%%%%%%%%%%%%%%%%%%%%%%%%%%%%%%%%%%%%%%%%%%%%%%%%%%%%%%

\paragraph{Retrieval and analysis of vibrational resonances}
We model the effective (contracted) nonlinear susceptibility as a coherent sum of resonant and non-resonant terms \cite{lambert2005,shen2016,pickering2022}:
\begin{equation}
\chi^{(2)}(\pm\omega_{\mathrm{IR}})
= \sum_{j=1}^{N} A_j e^{i\phi_j}\,
  \frac{1}{\omega_j-\omega_{\mathrm{IR}} \mp i\Gamma_j/2}
  + A_{\mathrm{nr}} e^{i\phi_{\mathrm{nr}}},
\end{equation}
where the upper `--' sign in the denominator corresponds to $\chi^{(2)}(+\omega_{\mathrm{IR}})$, i.e., SFG at $\omega_{\mathrm{VIS}}+\omega_{\mathrm{IR}}$, and the lower  `+' ) sign to $\chi^{(2)}(-\omega_{\mathrm{IR}})$ (DFG). For conciseness, since the VIS laser wavelength is fixed in all measurements, we note $\chi^{(2)}(\pm\omega_{\mathrm{IR}})=\chi^{(2)}(\omega : \omega_{\rm VIS},\pm\omega_{\mathrm{IR}})$. The power-normalized signals follow
\begin{equation}
{I}^{(2)}(\pm \omega_{\mathrm{IR}})\propto \big|\chi^{(2)}(\pm \omega_{\mathrm{IR}})\big|^{2}\,  \eta_{\rm overlap}(\omega_{\mathrm{VIS}}, \omega_{\mathrm{IR}}, \omega_{\mathrm{VIS
}}\pm \omega_{\mathrm{IR}})^2
\label{eq:intensity_main}
\end{equation}
where $\eta_{\rm overlap}^2$ is the plasmonic nonlinear mode overlap that depends on the spatial distributions of $\chi^{(2)}(\vec r)$ and of the the near-fields at all three frequencies involved \cite{hu2025}. It captures the field enhancement and colocalization of the modes: $1/\eta_{\rm overlap}^2$ is an effective nonlinear mode volume; the smaller it is, the larger is the nonlinear signal enhancement \cite{hu2025}.
The fact that the smooth envelope to the SFG and DFG intensities follows the IR absorption spectrum (see Fig.~\ref{fig:arrays}c–d) is consistent with $\eta_{\rm overlap}$ varying across the MIR resonance \cite{hu2025}. 

\begin{table}[h]
\centering
\begin{tabular}{ll}
\hline
\textbf{Symbol} & \textbf{Description} \\
\hline
$A_j$ & Amplitude of the $j$-th vibrational mode response \\
$\phi_j$ & Phase offset of the $j$-th mode \\
$\omega_j$ & Resonance frequency of the $j$-th mode (fixed) \\
$\Gamma_j$ & Linewidth of the $j$-th mode (fixed) \\
$A_{\mathrm{nr}}, \phi_{\mathrm{nr}}$ & Amplitude and phase of the nonresonant $\chi^{(2)}$ response \\
$\omega_{\mathrm{IR}}$ & Incident MIR frequency (experimentally controlled) \\
\hline
\end{tabular}
\caption{Physical interpretation of fitting parameters.}
\label{tab:Parameteres}
\end{table}

In Fig.~\ref{fig:broadband}, to evidence the BPhT vibrational resonances, we divided the DFG and SFG spectra by a smooth curve that accounts for the MIR resonant enhancement of the metasurface. Alternatively, 
to obtain a $\chi^{(2)}$ readout that is self-normalized and resilient to sample drift and laser power fluctuation, we can compute the ratio (Fig.~\ref{fig:broadband}a)
\[
R(\omega_{\mathrm{IR}})=\frac{\tilde{I}^{(2)}(+ \omega_{\mathrm{IR}})}{\tilde{I}^{(2)}(- \omega_{\mathrm{IR}})}.
\]
After factoring out the plasmonic MIR resonance, the spectra of Fig.~\ref{fig:broadband}a share a flat baseline due to the non-resonant electronic response, and exhibit sharp peaks, dips, or dispersive features at the vibrational resonances, sensitive to the phase difference $(\phi_j-\phi_{\mathrm{nr}})$. %A detailed discussion of the single–vibrational mode case is provided in the SI. 
We perform a {joint} fit of SFG, DFG, and $R(\omega_{\mathrm{IR}})$ using Bayesian optimisation to enforce cross–channel consistency and accelerate convergence \cite{martinez-cantin2014,wiechers2023}. A single, self–consistent set of $\chi^{(2)}$ parameters fits all three simultaneously acquired curves. This approach leverages the complementary phase sensitivity of SFG and DFG \cite{balos2022} and the baseline and power–fluctuation suppression inherent to $R(\omega_{\rm IR})$, which together reduce parameter degeneracy and tighten confidence intervals compared with indepedent fits of the three curves (see SI, section S5).

%%%%%%%%%%%%%%%%%%%%%%%%%%%%%%%%%%%%%%%%%%%%%%%%%%%%%%%%%%%%%%%%

    \begin{figure}[h] % the position of the figure, does it matter?
        \centering
        \includegraphics[width=0.5\linewidth]{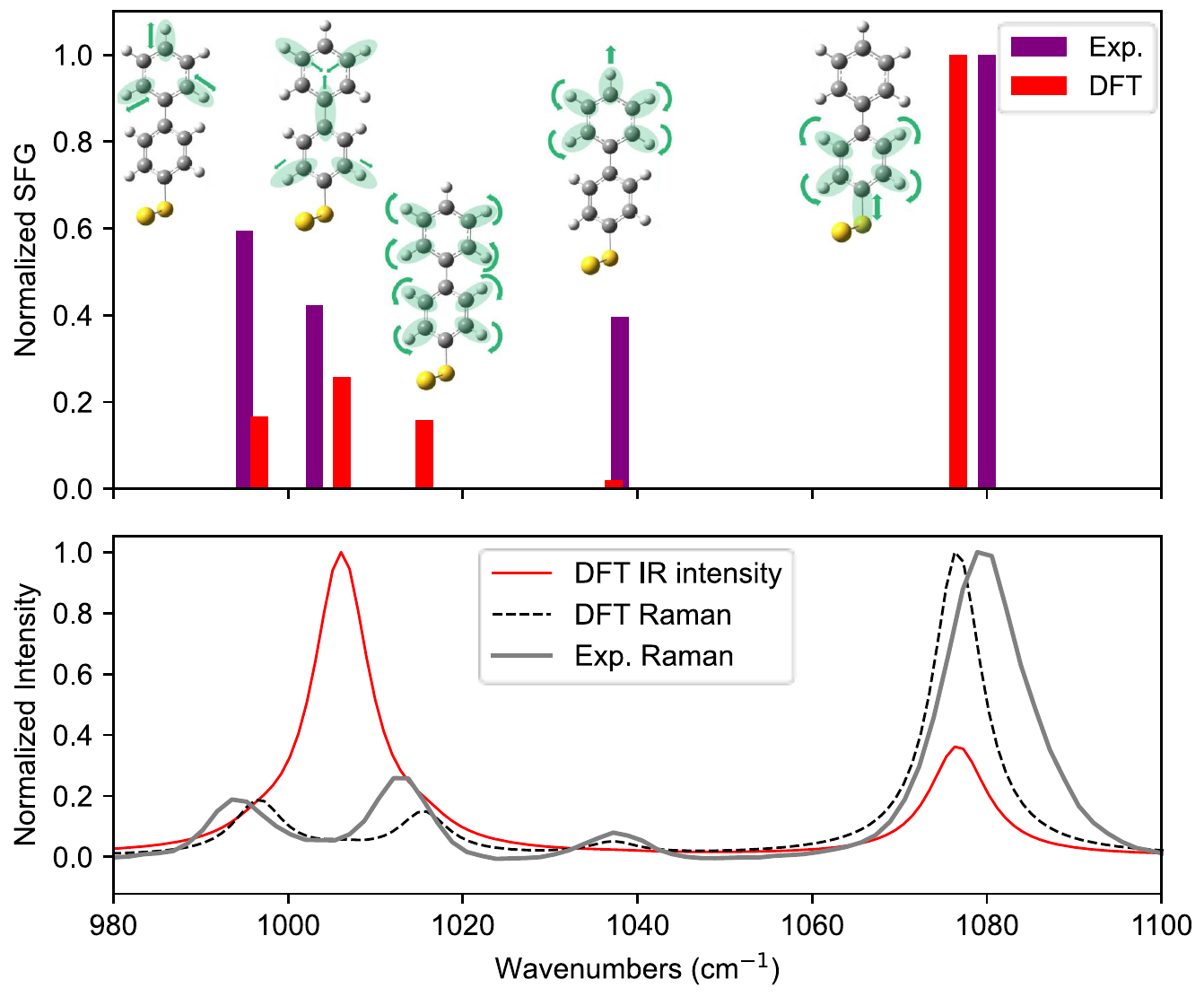}
        \caption{DFT calculations for an \emph{Au–S capped} molecule (thiol hydrogen replaced by a single Au atom to model chemisorption) provide a baseline for mode positions and for the expected sign of the resonant $\chi^{(2)}_{zzz}$ contribution (bar plot: red/blue denote opposite signs), together with the corresponding DFT Raman and IR spectra (broadened for readability). While positions broadly align, the measured amplitudes, line shapes, and even signs can deviate in a robust, cavity-specific manner, consistent with electromagnetic/chemical environmental effects in the nanogap that are not captured by the single-atom adsorption model.}
        \label{fig:DFT}
    \end{figure}
%%%%%%%%%%%%%%%%%%%%%%%%%%%%%%%%%%%%%%%%%%%%%%%%%%%%%%%%%%%%%%%%

%\paragraph{Zoomed ratiometric readout and comparison to a minimal adsorption model}
To illustrate the rich information contained in our spectra, we examine more closely in Fig.~\ref{fig:broadband}b the spectra from three other distinct NPoS across M-2 and M-3 metasurfaces, over the 970–1110~cm$^{-1}$ MIR range (more examples are shown in Figs. S5 and S6). %The ratio suppresses common-mode baselines from the shared MIR path and yields sub-wavenumber contrast. 
We reproducibly resolve a triplet near 1{,}000~cm$^{-1}$: in addition to the two Raman peaks at 995 and 1{,}012~cm$^{-1}$, coherent SFG and DFG clearly resolve an additional resonance at 1{,}003~cm$^{-1}$. This vibrational mode, weakly Raman active, must therefore present a large IR dipole. The variations among different spectra are most likely reflecting the non-uniform orientation of BPhT molecules across different samples and cavities \cite{azzam2002,kett2012, takahashi2025}, which opens up opportunities for future studies.

We now compare these observations to DFT calculations of a Au–S terminated BPhT molecule (thiol hydrogen replaced by a single Au atom; see Section~S6 of SI for methods). After application of a usual scaling factor, the predicted vibrational mode frequencies match well the experimental data. The SERS spectrum is closely reproduced by the DFT calculations. 
However, the DFT predictions of the resonant $\chi^{(2)}$ fail to reproduce key experimental observations; notably, they tend to overestimate the relative strengths of the modes at 1{,}012~cm$^{-1}$ and 1{,}080~cm$^{-1}$, they underestimate the $\chi^{(2)}$ response of the mode at 1{,}038~cm$^{-1}$, and they cannot explain the distribution of relative phases $\phi_j-\phi_{\mathrm{nr}}$. At the molecular level, $\phi_j$ can only take two values, $0$ or $\pi$, depending on how the Raman and IR dipole grow and shrink with normal mode displacement. But it appeared impossible to fit our data with such a constraint (as also recently reported in \cite{barbillon2025}).
This may indicate that electromagnetic and chemical environmental effects in the nanogap are not captured by the single-atom adsorption model. A microscopic assignment is beyond the scope of this work; possible mechanisms include chemical hybridization at the metal–molecule interface, charge-transfer electronic resonances, and lifetime modifications under enhanced MIR fields are considered \cite{metzger2019,verlekar2025}.

%%%%%%%%%%%%%%%%%%%%%%%%%%%%%%%%%%%%%%%%%%%%%%%%%%%%%%%%%%%%%%%%
    \begin{figure}[h]
        \centering
        \includegraphics[width=1\linewidth]{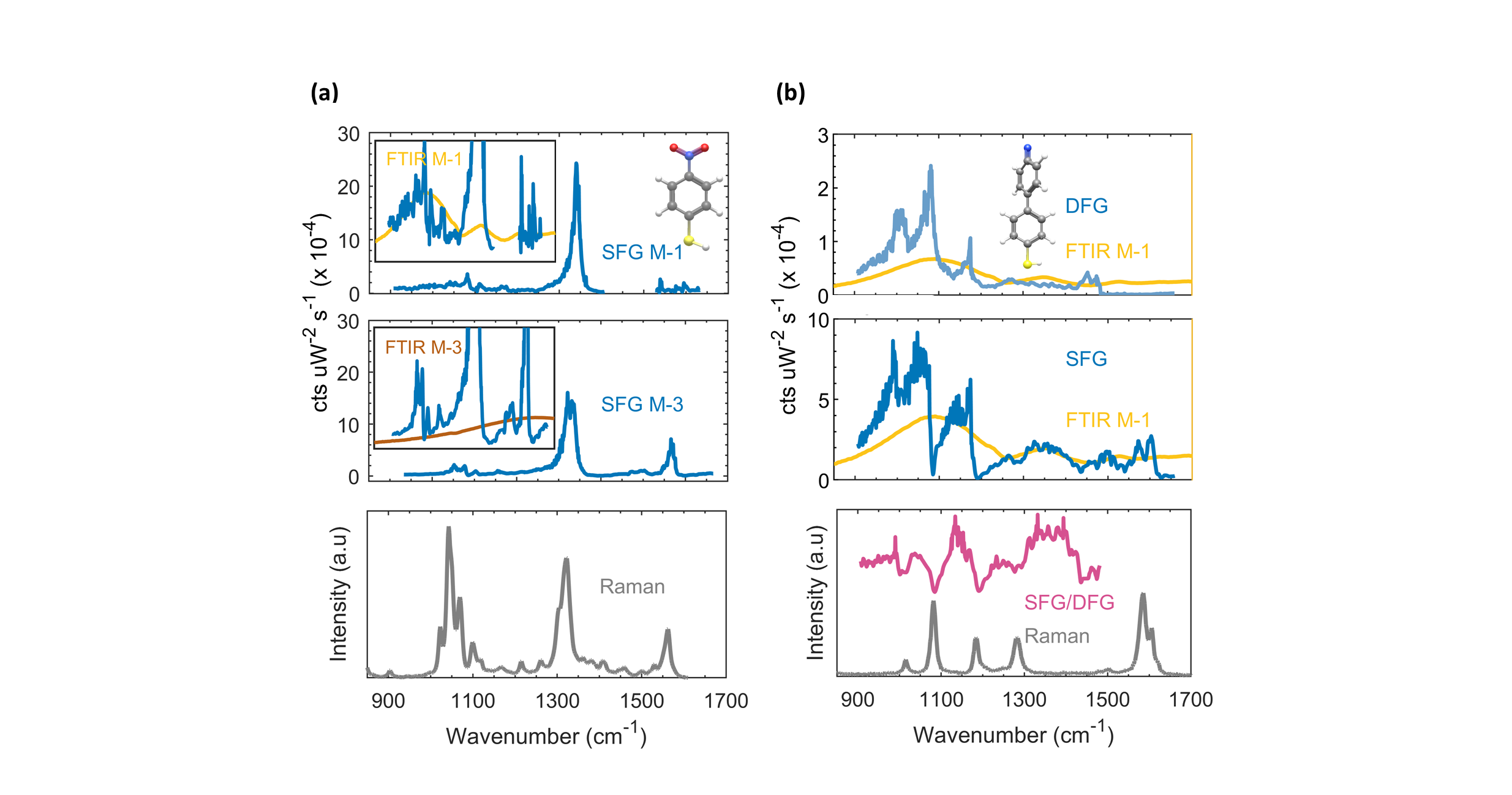}
        \caption{\textbf{(a) 4-NTP on two metasurfaces.} Top: SFG spectrum from a single NPoS on M-1. Inset: overlay with the M-1 FTIR spectrum, yellow line. Middle: SFG spectrum from a single NPoS on M-3. Inset: overlay with the M-3 FTIR spectrum, orange line. Bottom: corresponding SERS spectrum. For 4-NTP the VIS power was limited to 1\,\textmu W to avoid photochemistry. \textbf{(b) cn-BPhT on M-1.} Power-normalized DFG (top) and SFG (middle) from a single NPoS, overlaid with the M-1 FTIR spectrum (yellow); bottom: SFG/DFG ratio and co-recorded SERS. 
        }
        \label{fig:ntp}
    \end{figure}
%%%%%%%%%%%%%%%%%%%%%%%%%%%%%%%%%%%%%%%%%%%%%%%%%%%%%%%%%%%%%%%%

\paragraph{Applicability to other molecules}
To test the generality of our platform beyond BPhT, we measured single–NPoS spectra for 4'-cyanobiphenyl-4-thiol (cn-BPhT) and 4-nitrothiophenol (4-NTP) molecules embedded in the gap (Fig.~\ref{fig:ntp}). In both cases, we show results obtained by functionalising the nanoparticles instead of forming a SAM on the substrate (see Sec. S2B in the SI). %Source–structure co-programming selects the MIR sub-band of interest without changing alignment: the metasurface fixes the MIR envelope, while the same cw VIS/MIR beams and detection are used throughout.
For {4-NTP} (Fig.~\ref{fig:ntp}a), photochemical reactivity imposed limiting the VIS power to $\sim$1\,\textmu W \cite{huang2010a,zhang2015b,sarhan2019,jeong2024,yao2024}. Under these conditions, when using the M-1 metasurface whose MIR response weakens above 1{,}400\,cm$^{-1}$, the SFG falls below the threshold for reliable analysis in the range from 1{,}500–1{,}600\,cm$^{-1}$. Switching to the the M-3 metasurface, which targets the 1{,}600\,cm$^{-1}$ band, restores a stronger response in this range. This illustrates the core design rule: by selecting the MIR band (QCL chip) and the slit geometry, we can tune the MIR enhancement to the molecular modes of interest and measure strong nonlinear signals even when the VIS power must be kept very low.
Compared with BPhT, 4-NTP exhibits a much larger resonant-to-nonresonant $\chi^{(2)}$ ratio when $\omega_{\mathrm{IR}}$ addresses the IR- and Raman-active modes near 1{,}340\,cm$^{-1}$ (NO$_2$ stretch) and $\sim$1{,}580\,cm$^{-1}$ (C=C ring stretch), consistent with reports of strong IR dipole–dipole interactions in self-assembled monolayers \cite{gray2021,mueller2022,bell2025,shin2023}. 

For {cn-BPhT} measured on the M-1 metasurface (Fig.~\ref{fig:ntp}b), the power-normalized SFG and DFG display pronounced dispersive line shapes, explained above as the interference between resonant (vibrational) and non-resonant contributions to the effective $\chi^{(2)}$. In this molecule, the two terms are comparable in magnitude, so both constructive and destructive features emerge clearly. Because the DFG is too weak close to $\sim$1600\,cm$^{-1}$, the ratiometric analysis $R(\omega)$ is evaluated only below 1500\,cm$^{-1}$. Within this window, the line shapes are reproducible across cavities. Compared with BPhT, which shares the biphenyl scaffold, cn-BPhT exhibits a different balance between resonant and non-resonant $\chi^{(2)}$ under identical cw conditions. We refrain from assigning a microscopic origin, although differences in electronic structure (e.g., the \textit{para}-cyano substituent) could plausibly influence the contribution of charge-transfer states to the non-resonant $\chi^{(2)}$ of the molecule-metal interface. This example illustrates that closely related molecules with similar SERS spectra can yield very distinct nanocavity SFG/DFG spectra. %The single-cavity readout makes these differences explicit, and metasurface selection simply targets the MIR sub-band of interest.

To position our continuous-wave SFG/DFG nanocavity platform within the MIR spectroscopy landscape, we benchmark it against representative methods using literature reports (Table~\ref{tab:benchmark}). The axes we emphasise are: (i) \emph{coherency} (phase–sensitive, background–rejecting upconversion); (ii) \emph{simultaneity} (SFG and DFG from the \emph{same} nano-site, enabling ratiometric $R(\omega)$); (iii) \emph{monolithic operation} with a tunable MIR source (no ultrafast laser, no phase matching constraints ); (iv) \emph{label-free specificity}; and (v) \emph{sensitivity} to few molecules.
This comparison clarifies that our platform is advantageous when {few-molecule sensitivity} and {compact, alignment-light hardware} are required together with {sub-wavenumber, laser-defined resolution} and {simultaneous} SFG/DFG for robust lineshape analysis under ambient conditions. %In contrast, free-space pulsed SFG excels at resolving ultrafast vibrational dynamics, nano-FTIR offers nanoscale mapping with an AFM but is incoherent and instrumentally heavier, and O-PTIR systems deliver turnkey photothermal imaging on the same hardware class; in default mode they do not report coherent upconversion, yet with suitable settings (and a metasurface sample) the optics and alignment can run our vSFG/vDFG workflow. Rather than ranking methods, Table~\ref{tab:benchmark} maps complementary operating regimes and highlights where a cw, single-nanocavity, co-registered readout is the practical choice.

%%%%%%%%%%%%%%%%%%%%%%%%%%%%%%%%%%%%%%%%%%%%%%%%%%%%%%%%%%%%%%%%%%%%%%%%
% Another suggestion 
\begin{table}[]
%\resizebox{\textwidth}{!}{%
\begin{tabular}{|l|c|c|c|c|c|}%{|l|l|l|l|l|l|}
\hline
Method & Coherent & Simult. SFG+DFG & Standalone MIR source & Label-free$^a$ & Number of molecules  \\ \hline
This work                       &  Yes  & Yes & Yes & Yes & $<10^3$   \\
Nanocavity pulsed SFG \cite{bell2025} &  Yes  &  Possible  &  No  & Yes & $<10^3$  \\
Free-space pulsed SFG \cite{shen2016} &  Yes &  Possible but complex  & No & Yes & typ. $>10^9$ \\
O-PTIR \cite{anderson2023,jia2024} & No & (NA) & Yes & Yes & $<10^3$    \\
BonFIRE (VEF) \cite{wang2023} &  No  &  (NA)  &  No  & No & single \\  
MIRVAL (VEF) \cite{chikkaraddy2023} &  No  &  (NA)  &  No  & No & $^c$ $1-10^3$ \\
SEIRA on a SAM \cite{aouani2013} &  No  &  (NA)  &  Yes  & Yes & typ. $>10^6$ \\
nano-FTIR$^b$ \cite{nishida2024} & No & (NA)  & Yes & Yes & $<10^3$ \\ \hline
\end{tabular}%
\caption{
\textbf{Contextual comparison of ambient, broadband MIR spectroscopies.}
Columns indicate whether a method is coherent (frequency conversion), whether {simultaneous} vSFG and vDFG can be recorded from the same site, whether a standalone tunable MIR source (e.g., cw QCL) suffices, whether fluorescent labels are necessary or not (label-free), and the typical number of molecules probed that are probed. Abbreviations: O-PTIR, optical photothermal infrared; VEF, vibrationally encoded fluorescence; SEIRA, surface-enhanced infrared absorption. $^{a}$“Label-free” denotes no fluorescent reporter. $^{b}$Nano-FTIR requires an AFM-based s-SNOM apparatus. $^{c}$Single-molecule sensitivity reported under stochastic picocavity formation. Reported molecule counts are order-of-magnitude estimates and depend on hotspot volume and coupling.}
%}
\label{tab:benchmark}
\end{table}
%%%%%%%%%%%%%%%%%%%%%%%%%%%%%%%%%%%%%%%%%%%%%%%%%%%%%%%%%%%%%%%%%%%%%%%%

%%%%%%%%%%%%%%%%%%%%%%%%%%%%%%%%%%%%%%%%%%%%%%%%%%%%%%%%%%%%%%%%%%%%%%%

\paragraph{Discussion } 
All data presented above establish a reliable plasmonic platform combined with a simple laser technique to investigate the combined MIR-VIS nonlinear response of metallic nanogaps. Such platforms are taking central roles in a growing number of research and application fields, such as photo-catalysis \cite{wy2022}, molecular and nano opto-electronics \cite{farmakidis2019,amirtharaj2024}, direct photodetection \cite{shi2011}, frequency upconversion \cite{roelli2020,xomalis2021,koczor-benda2022,koczor-benda2025}, nanoscale thermal transport \cite{mosso2019,guo2022}, polaritonics \cite{sanchez-barquilla2022}, vibrational strong coupling \cite{dayal2021}, 2D material opto-electronics \cite{yao2014a}, biosensors \cite{yang2020}, chemical fingerprinting \cite{kim2025}, single-molecule spectroscopy \cite{g.etchegoin2008,taniguchi2017}, intra-molecular dynamics \cite{wilcken2023}, etc. Plasmonic enhancement is a useful capability for applications requiring high throughput, such as hyperspectral far-field \cite{prater2024,fang2024,niemann2024} or near-field \cite{schnell2010,sakurai2025,roelli2025, takahashi2025} MIR and SFG imaging. 

Several observations call for future studies beyond the present work. These include the deviations between DFT prediction and experimental SFG spectra (while the SERS is in good agreement), the marked molecular dependence of the relative strength of resonant (vibrational) vs. non-resonant nonlinear response, and the exact origin of the four-wave mixing signal. To shed light on these phenomena, a first parameter to be explored is the VIS wavelength dependence of the various signals. Tuning the VIS laser will clarify how the plasmonic resonance and the electronic structure of the metal can alter the nonlinear response of a nanocavity \cite{dreesen2002,dalstein2018}. Second, the possible contribution of a charge-transfer state to the nonlinear response (similar to that reported for SERS  \cite{arenas2002,park2010,zhang2015a}) can be clarified by systematically modifying the molecules with electron-donating/withdrawing substituents \cite{imahori2021} or applying static electric fields \cite{kirch2024}. The latter approach opens a new dimension in the parameter space, as electric fields can also break inversion symmetry and thereby enhance the second-order response \cite{fischer2003,mollica2020,zheng2023,hogue2023}.

%%%%%%%%%%%%%%%%%%%%%%%%%%%%%%%%%%%%%%%%%%%%
\paragraph{Conclusion and perspectives}
We upgraded coherent vibrational spectroscopy from an ultrafast, phase-matched ensemble measurement into a tunable continuous-wave, single-nanocavity, multi-channel platform. A single MIR frequency sweep on one nanocavity yields four complementary spectroscopic channels: SFG, DFG, FWM, and Raman scattering. This capability is useful for disentangling resonant and non-resonant contributions, probing charge-transfer–assisted nonlinearities, and characterizing surface nonlinearities of the nanostructure \cite{li2017c,vabishchevich2023,roelli2024,hanke2009,palomba2009,thyagarajan2012,butet2015,metzger2015,meier2023}. The nonlinear signals are acquired with a QCL-defined resolution better than $1$\,cm$^{-1}$, without geometric phase matching constraints, and under ambient conditions. Compared to conventional SFG spectroscopy, our approach bridges coherent nonlinear spectroscopy with chip-integrated chemical analysis under ordinary environments. It offers higher spectral resolution, much lower molecular detection limits, compact implementation, and direct compatibility with commercial MIR-Raman microscopes \cite{prater2020,prater2024,jia2024}.

Our work establishes a new path for the study of vibrational phenomena at the single-molecule limit, which can be reached using host–guest architectures \cite{kim2018} or DNA-origami trapping \cite{hubner2019,verlekar2025}. This includes the study of mode-specific, MIR-induced chemical reactions \cite{stensitzki2018,pannir-sivajothi2025}, for which the SERS functionality built into our platform is a powerful way to track reactivity with single-molecule sensitivity \cite{li2020}.
More generally, we foresee that our approach will enable new insights into intra-molecular \cite{rubtsov2009,kasyanenko2011,cohn2025} and inter-molecular \cite{xiang2020,hirschmann2024,zheng2025} vibrational relaxation pathways in plasmonic environments. 
Specifically, entering the single-molecule regime will make it possible to probe vibrational anharmonicity, which is predicted to yield non-classical photon statistics in SFG \cite{kalarde2025}, opening applications in quantum-enhanced spectroscopy. The cavity-enhanced MIR excitation has also been predicted to enable Floquet engineering of molecular dynamics \cite{reitz2020}, resulting in a laser-tunable strength and linewidth of some vibrational sidebands.  Such coherent signatures go beyond the reach of recently developed single-molecule IR methods based on incoherent readouts, such as vibrationally-encoded fluorescence \cite{xiong2019,whaley-mayda2021,calvin2023,chikkaraddy2023,wang2023,wang2024,lee2025}.

\section*{Acknowledgments}
F. C. thanks Roberto A. Boto for insightful exchanges about the DFT modelling. Z. X. and C. G. thank Matthias Godejohann for valuable assistance with the MIR laser, and the EPFL Cmi and IPHYS staff for technical support, and Arnaud Jollivet for earlier contributions to the setup. This research received funding from the Swiss National Science Foundation (SNSF project No. 214993) and the European Union's Horizon 2020 research and innovation program under Grant Agreement No. 820196 (ERC CoG `QTONE')
\bibliography{Broadband_SFG}

% Numbering setup for SI
\renewcommand{\thesection}{S\arabic{section}}
\renewcommand{\thefigure}{S\arabic{figure}}
\renewcommand{\thetable}{S\arabic{table}}
\setcounter{section}{0}
\setcounter{figure}{0}
\setcounter{table}{0}
% Fix references
\makeatletter
\renewcommand{\p@subsection}{}
\renewcommand{\p@subsubsection}{}
\makeatother

%%%%%%%%%%%%%%%%%%%%%%%%%%%%%%%%%%%%%%%%%%%%%%%%%%%%%%%%%%%%%%%%%%%%%%%%%%
\section{Supplementary Figures}

\begin{figure}[H]
\centering
\includegraphics[width=1\linewidth]{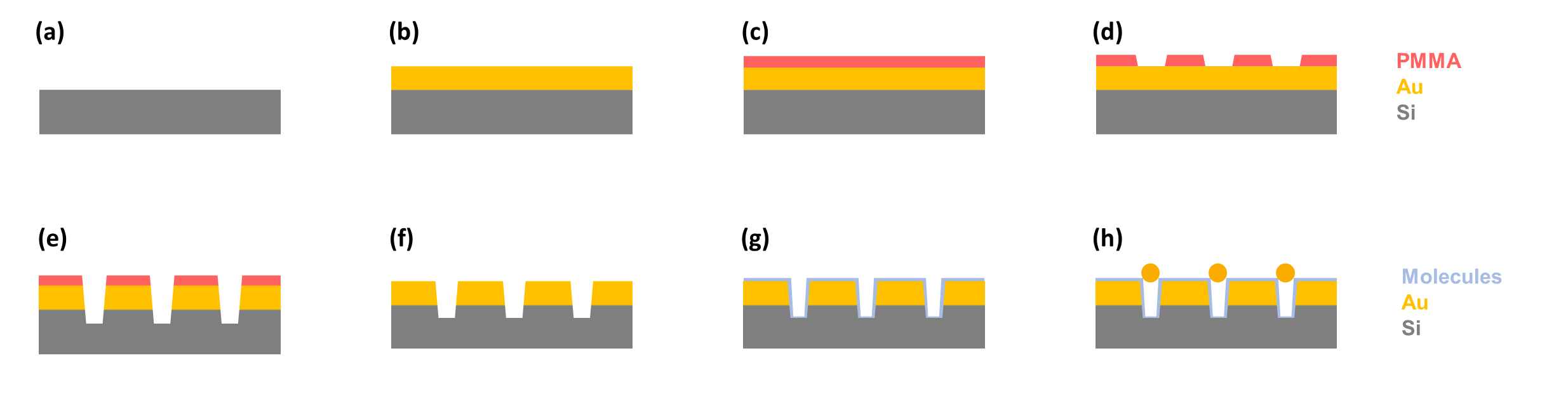}
\caption{
Schematic illustration of the nanostructure fabrication and functionalization process.
(a) Double-polished silicon (Si) wafer.
(b) Thermal evaporation of a 5-nm chromium (Cr) adhesion layer and 150-nm gold (Au) layer.
(c) Spin-coating of 950A4 PMMA resist with 2000 rpm.
(d) Electron-beam lithography (EBL) to define nanoslit patterns.
(e) Ion-beam etching to create V-shaped trenches.
(f) Removal of the remaining PMMA resist.
(g) Formation of a self-assembled monolayer (SAM).
(h) Deposition of gold nanoparticles, some of them captured in the nanoslits.
}

\end{figure}

\begin{figure}[H] 
\centering
\includegraphics[width=0.7\linewidth]{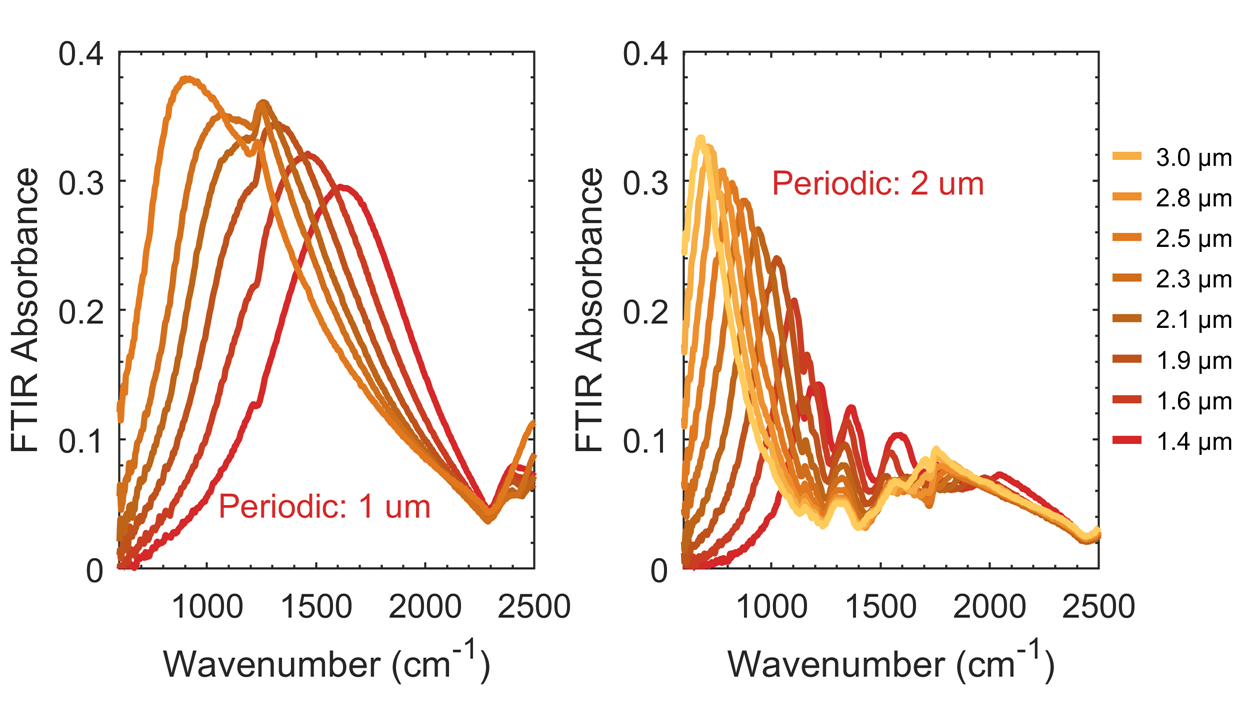}
\caption{FTIR absorbance spectra of nanostructures with varying slit lengths (1.4–3.0~µm), for two periodicities: 1~µm (left) and 2~µm (right). 
By tuning the slit length, the plasmonic resonance is shifted across the mid-IR region, enabling spectral matching with specific molecular vibrational bands.
Structures used in the main experiments were selected from these sets to optimize coupling with target molecular modes.
}
\end{figure}

\begin{figure}[H]
\centering
\includegraphics[width=0.8\linewidth]{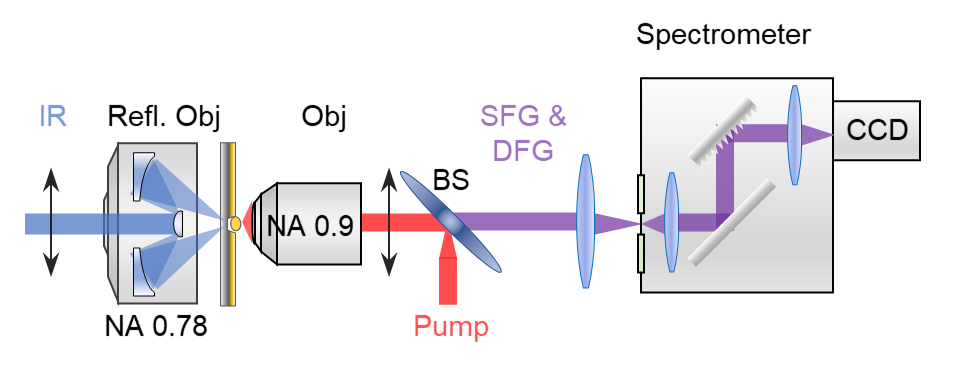}
\caption{
Schematic of the experimental setup for sum- and difference-frequency generation (SFG/DFG) spectroscopy. 
A continuous-wave, monochromatic MIR beam (continuously tunable from 860 to 1670~cm$^{-1}$, MIRcat QCL) is focused onto the sample through a reflective objective (NA~0.78), while a counter-propagating 785~nm pump beam is delivered via a refractive objective (NA~0.9). 
The generated Raman and nonlinear (SFG and DFG) signals are collected by the same objective and directed to a spectrometer (Kymera 193i or Shamrock 750) via a beamsplitter (BS). Notch filters are used to suppress the residual pump. 
The sample was mounted on a piezo-controlled XYZ stage and measured under ambient conditions.
}

\end{figure}

\begin{figure}[H] 
\centering
\includegraphics[width=0.4\linewidth]{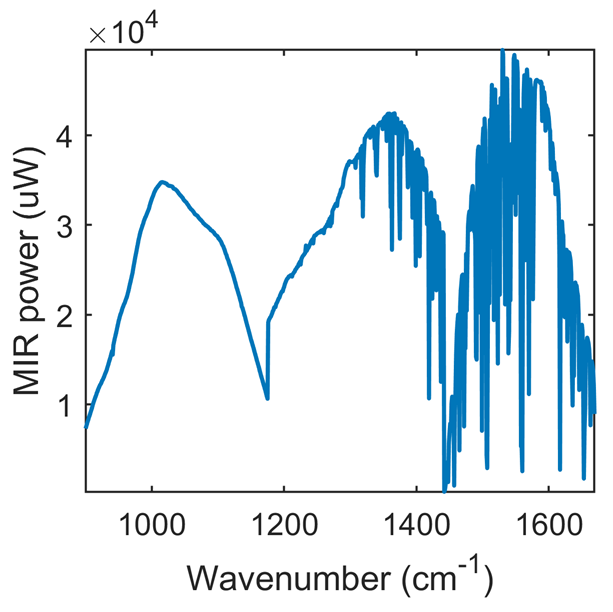}
\caption{Measured MIR power at the entrance of the reflective objective as a function of wavenumber, recorded using a Thorlabs PM100D power meter equipped with an S470C thermopile sensor, positioned before the focusing objective. 
This measurement reflects the actual MIR power incident on the sample under the stepwise acquisition configuration. 
The spectral envelope corresponds to the tuning profile of the MIRcat QCL system (860–1670~cm$^{-1}$), while sharp absorption lines above 1300~cm$^{-1}$ are due to water vapour and other gases.
}
\end{figure}

\begin{figure}[H] 
\centering
\includegraphics[width=1\linewidth]{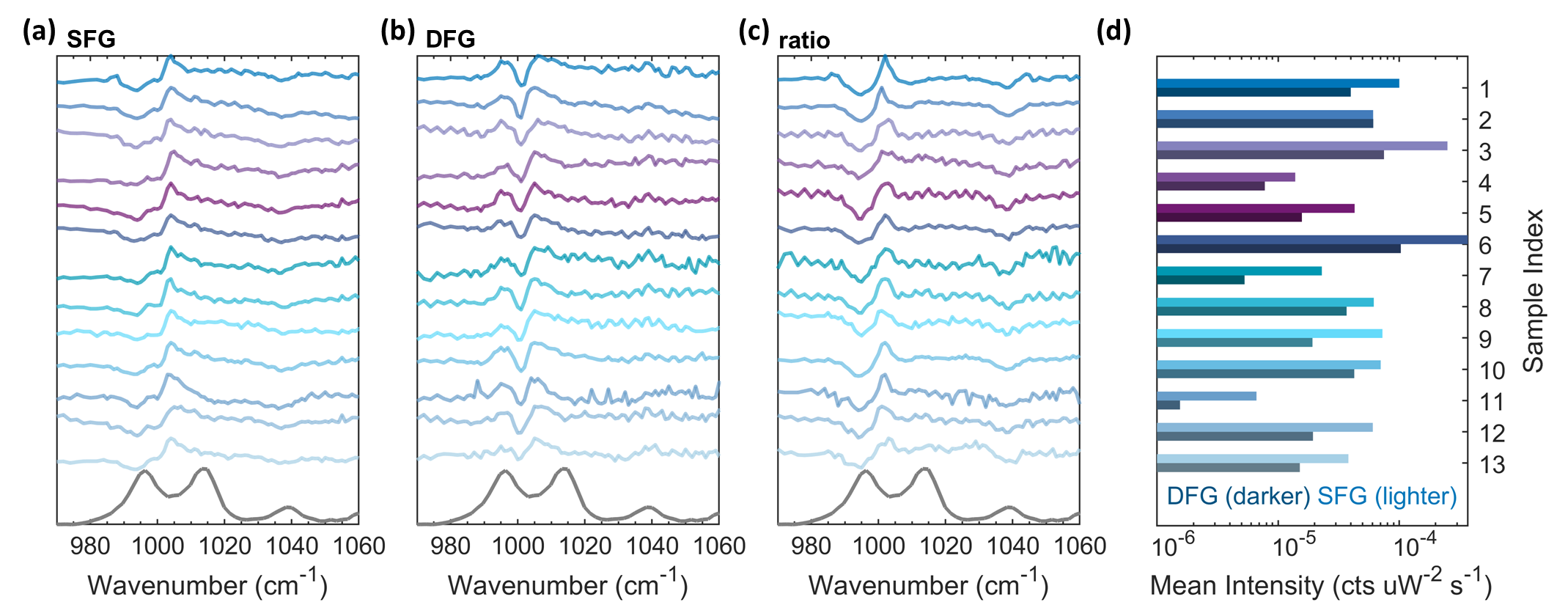}
\caption{
Characterization of all tested NPoS on a same chip. 
(a)~SFG spectra, (b)~DFG spectra, and (c)~their SFG/DFG ratio for 13 randomly selected locations on a single device, illustrating variation in signal strength and line shape, while the main vibrational features are robust. 
The bottom grey trace in each plot corresponds to the reference SERS spectrum of the target molecule. 
(d)~Quantitative comparison of the mean intensities of SFG (light) and DFG (dark) signals for each sample, normalized by input powers. 
All measurement results were processed with the data processing method described in \ref{sec:processing}}
\end{figure}

\begin{figure}[H] 
\centering
\includegraphics[width=0.6\linewidth]{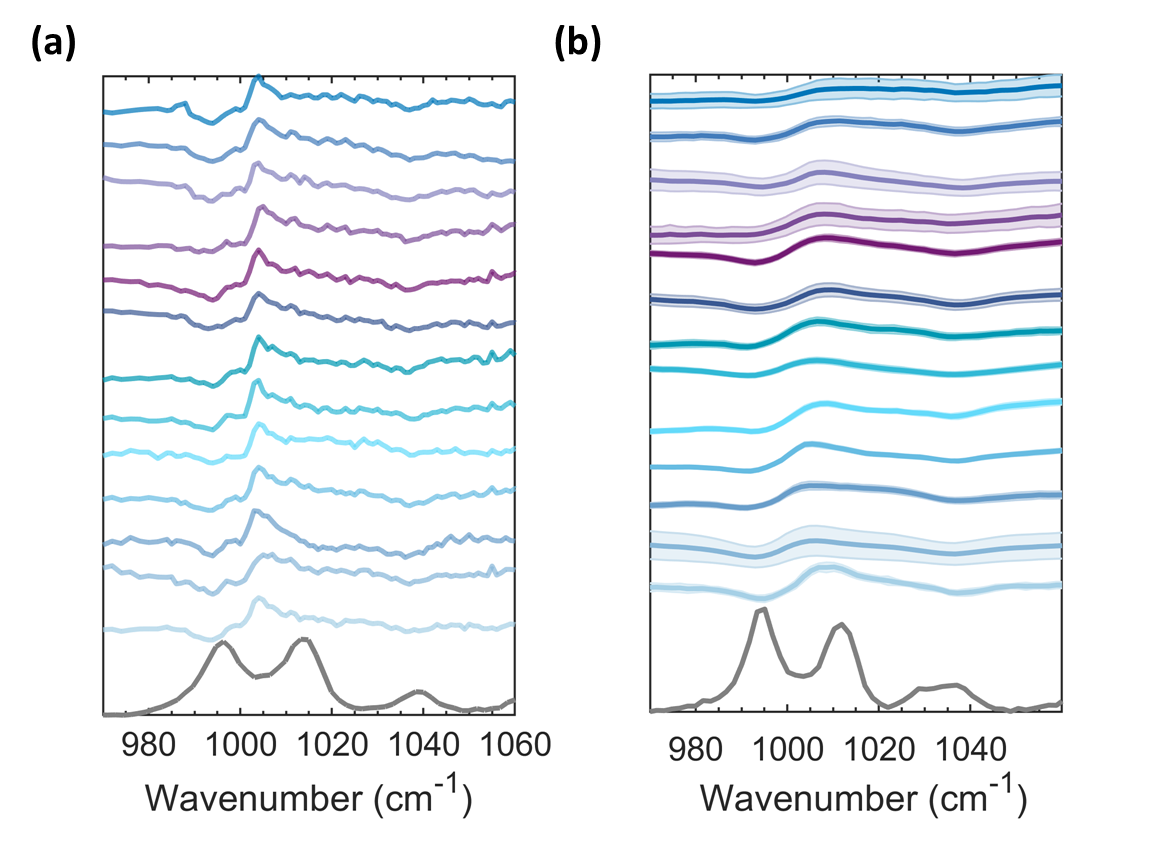}
\caption{
Comparison between stepwise (a) and fast-scan (b) acquisition methods of SFG spectra. Curves with identical colors correspond to the same NPoS structure. While DFG signals are only detectable in stepwise scans due to the strong spontaneous Stokes Raman scattering, the SFG spectral features are reproducible in both methods, demonstrating the qualitative utility of fast acquisition.
}
\end{figure}

%\begin{figure}[H] \centering\includegraphics[width=0.6\linewidth]{S 1080.png}\end{figure}

\begin{table}[h]
\centering
\begin{tabular}{lccc}
\hline
Metasurface & Molecule & $\mathrm{median}\,\eta_2$ & 1\,s SNR$_{\mathrm{peak}}$ \\
\hline
M-2 & BPhT     & $8.5\times10^{-2}$\,(IQR $4.2$–$10.1\times10^{-2}$) & $14$ (IQR $10$–$22$) \\
M-2 & 4-NTP@1340 & $1.7\times10^{-1}$\,(IQR $1.1$–$2.2\times10^{-1}$) & $>20$ \\
M-3 & BPhT     & $1.3\times10^{-3}$\,(IQR $0.7$–$2.0\times10^{-3}$) & $9$  (IQR $7$–$12$) \\
M-3 & cn-BPhT  & $2.1\times10^{-3}$\,(IQR $1.3$–$3.4\times10^{-3}$) & $11$ (IQR $8$–$15$) \\
\hline
\end{tabular}
\caption{Power-normalised peak vSFG efficiency $\eta_2$ and 1\,s peak SNR across single cavities. Cross-design values are also reported as $\tilde{\eta}_2=\eta_2/\mathrm{FTIR}_{\max}$ in Table~S\# to mitigate coupling differences.}
\end{table}

%%%%%%%%%%%%%%%%%%%%%%%%%%%%%%%%%%%%%%%%%%%%%%%%%%%%%%%%%%%%%%%%%%%%
\section{Fabrication of Nanostructured Devices}
\label{sec:fabrication}
\subsection{Slit fabrication}
All devices were fabricated using 380-nm-thick, double-polished silicon wafers. A 5-nm chromium (Cr) adhesion layer and a 150-nm gold (Au) layer were thermally evaporated onto the wafer surface at a controlled deposition rate of 0.5~nm/s.
Electron-beam lithography (EBL) was used to define nanoslit patterns. Wafers were spin-coated with PMMA 950 A4 at 2000 rpm and baked to ensure uniform coverage. Across all designs, the nanoslit length along the slit axis was fixed at 2~\textmu m, while the slit widths is 140~nm. The periodicity along the slit axis was kept constant, whereas the periodicity perpendicular to the slit was varied to tune the plasmonic resonance. The patterns were transferred into the Au layer via collimated ion-beam etching at a -10$^\circ$ angle to form V-shaped trenches, followed by a -70$^\circ$ etch to remove fencing at the slit edges. Specific slit widths and perpendicular periodicities for each design are listed in Table~\ref{tab:slit_designs}.

\begin{table}[ht]
\centering
\caption{Nanoslit geometry parameters for each EBL design. The slit length and parallel periodicity were fixed for all designs.}
\label{tab:slit_designs}
\begin{tabular}{lccc}
\hline
\textbf{Design} & \textbf{Slit Length (\textmu m)}& \textbf{Parallel Periodicity (\textmu m)} & \textbf{Perpendicular Periodicity (\textmu m)} \\
\hline
Slit-1 & 1.8
& 1.0& 2\\
Slit-2 & 1.4
& 1.0& 1\\
Slit-3 & 1.9& 1.0& 2\\

\hline
\end{tabular}
\end{table}

The wafers were diced into 1.5~cm~$\times$~1.5~cm chips. Remaining PMMA resist was removed using Remover 1165 (NMP) for \textgreater6~hours, rinsed with DI water, and dried with nitrogen. Piraniha solution treatment (5~minutes) ensured organic residue removal.

\subsection{Molecules and Nanoparticle Functionalization}
Surface-functionalization was carried out using 1~mM ethanolic solutions of molecules (BPhT, cn-BPhT, 4-NTP, Sigma-Aldrich) for 24~hours to form SAMs. Excess molecules were removed by ethanol rinsing and nitrogen drying. 
Then, 150~nm gold nanoparticles (Nanopartz, OD 2) in ethanol were deposited on the nanoslit regions. After 1~minute incubation, the sample was dried with nitrogen, rinsed with DI water, and dried again. Nanoparticles were randomly distributed on the nanoslits and the surrounding film.

We also noticed that it is also possible to perform experiments with functionalized nanoparticles instead of forming a monolayer on the metasurfaces. We tested functionalization of gold nanoparticles with cn-BPhT and 4-NTP using the following method. Citrate-capped gold nanoparticles ( 2.5 pM in 100 $\mu$L of water) were mixed with 1 mM ethanol solutions of cn-BPhT or 4-NTP (20 $\mu$L) and sonicated for 60 minutes at room temperature, followed by a 24-hour incubation period. For these molecules, this corresponds to $\sim 250\times$ the amount required for a close-packed monolayer on a 150-nm particle ($3.2\times 10^5$ molecules per particle, assuming 0.22 nm$^2$ per molecule). The samples were purified by centrifugation at 150,000 rpm for 10 minutes , followed by removal of the supernatant and redispersion in 1.0 mL of Milli-Q water. In the second cycle, 1.0 mL of ethanol was used to remove unbound thiols, after which the supernatant was again discarded. Finally, the aggregated nanoparticles were redispersed in 1.0 mL of ethanol. The final sample solutions were stored at 4 $^\circ$C in the dark. Finally, we deposited these functionalized nanoparticles on the slits, instead of the citrate-capped ones.

%%%%%%%%%%%%%%%%%%%%%%%%%%%%%%%%%%%%%%%%%%%%%%%%%%%%%%%%%%%%%%%%%%%%
\section{Additional Experimental Details}

\subsection{Stepwise CW Acquisition (Main Method)} \label{sec:step}

This section provides additional details on the stepwise MIR acquisition protocol used to generate the data presented in Fig.~2 and related figures.
The primary experimental method employed a fully automated stepwise acquisition scheme using a MIRcat QCL laser (Daylight Solutions). The laser was programmed to scan from 800 to 1670~cm$^{-1}$ in 1~cm$^{-1}$ steps. For each MIR wavelength, the following procedure was executed:

\begin{enumerate}
  \item The QCL was tuned to the target wavenumber and allowed to settle for 200–500~ms to ensure spectral and power stability.
  \item After the delay, a TTL trigger signal was sent from the control computer to initiate spectrum acquisition on the spectrometer.
  \item The spectrometer collected a Raman/SFG/DFG spectrum with an exposure time of 1 to 2~s, depending on signal intensity.
  \item Once acquisition was completed, the system advanced to the next MIR wavelength and repeated the process.
\end{enumerate}

The MIR output power could be measured after each spectrum if desired. However, as the laser output was verified to be stable over time, power calibration was typically performed only once every 1 to 2 months. The entire stepwise scan required approximately 40 minutes to complete and yielded high-resolution, high-SNR spectra suitable for both SFG and DFG analysis.

%%%%%%%%%%%%%%%%%%%%%%%%%%%%%%%%%%%%%%%%%%%%%%%%%%%%%%%%%%%%%%%%%%%%
\subsection{Alternative Acquisition Method: Fast MIR Scan} \label{sec:fast}

In addition to the stepwise CW acquisition described above, a faster experimental approach was employed for selected samples. Here, the MIRcat laser was operated in a rapid continuous scanning mode at 1800~cm$^{-1}$/s, with each spectrum acquired under 30~s exposure time, corresponding to hundreds of complete frequency sweeps. A reference Raman spectrum (without MIR excitation) was collected under identical conditions.

During data processing, Raman fluctuation compensation was first applied to all spectra using the same method as described below in the Data Processing methods. The reference Raman spectrum was then subtracted to isolate the nonlinear response. Finally, MIR power compensation was performed using recorded power profiles.

This method allows faster acquisition of SFG signals compared to stepwise measurements. While it is not suitable for resolving the DFG spectrum due to strong spontaneous Raman scattering on the Stokes side, it yields good results on the SFG (anti-Stokes) side and allows for rapid screening of vibrational resonances and broadband SFG response.

%%%%%%%%%%%%%%%%%%%%%%%%%%%%%%%%%%%%%%%%%%%%%%%%%%%%%%%%%%%%%%%%%%%%
\section{Data Processing method} \label{sec:processing} 

All spectra were recorded using a CW stepping mode setup. Raw data were processed through the following workflow:

\begin{enumerate}
    \item \textbf{Spectral Alignment:} \\
    The wavelength axis was calibrated by identifying the anti-Stokes and Stokes peaks and applying a small wavelength offset (within $-5$~nm to $+5$~nm) to align the corresponding wavenumber peaks symmetrically after conversion. The optimal offset was selected by randomly sampling within the allowed range and minimizing the mismatch between the two peak positions. This correction accounts for possible spectral shifts due to instrumental factors such as grating changes, slit width adjustments, laser frequency drift, etc.

    \item \textbf{Baseline Correction:} \\
    Each spectrum was baseline corrected using a modified version of a public domain polynomial fitting algorithm \footnote{{https://www.mathworks.com/matlabcentral/fileexchange/69649-raman-spectrum-baseline-removal}}. To improve numerical stability and robustness, we adjusted the fitting window to a fixed 33-point span (typically third-order polynomial) and occasionally to 9 points when required by sharp spectral structures. Spectra showing abnormal scaling behavior (outside the 0.7–1.5 range relative to the reference) were reprocessed with fallback parameters or excluded from the analysis.

    \item \textbf{Fluctuation Compensation:} \\
    To mitigate slow laser power fluctuations and sample drift, each spectrum was rescaled relative to a reference using selected stable regions in both anti-Stokes and Stokes domains (e.g., 700--850~cm$^{-1}$, 1120--1411~cm$^{-1}$, etc.).

    \item \textbf{Raman Reference Subtraction:} \\
    A representative Raman signal was reconstructed from selected regions of early and late spectra, combined into a reference signal (RamanF). This reference was subtracted from each spectrum to isolate the nonlinear signal.

    \item \textbf{Peak Detection and Integration:} \\
    SFG and DFG regions were analyzed independently. For each spectrum, Gaussian fitting was applied within user-defined pixel windows corresponding to the expected SFG and DFG features. These windows were selected based on prior calibration and adjusted as needed depending on the experimental configuration (e.g., grating or center wavelength). From the fitted peaks, both peak positions and integrated intensities were extracted for further analysis.

    \item \textbf{Outlier Removal and Denoising:} \\
    Spectra with power values below a given threshold or poor fitting results were excluded. Optional wavelet denoising (e.g., level 2--3) was applied to reduce noise in peak integrals.

    \item \textbf{Normalization and Output:} \\
    Final SFG and DFG integrals were normalized and plotted against MIR excitation wavenumber. Data were exported in csv format for quantitative analysis.
\end{enumerate}

%%%%%%%%%%%%%%%%%%%%%%%%%%%%%%%%%%%%%%%%%%%%%%%%%%%%%%%%%%%%%%%%%%%%
\section{Model-Based Spectral Fitting}\label{sec:fitting}

The fitting parameters used in Fig.2 is shown in Table~\ref{tab:fitparams}. The best-fit values were obtained from Bayesian optimization as described in the main text, using fixed resonance positions and damping factors. Each parameter was constrained within physically reasonable bounds derived from spectral knowledge.

\begin{table}[ht]
\centering
\caption{
Best-fit model parameters for the Slit-2 sample used in Fig.~2. Resonant frequencies \(w_i\), damping factors \(\gamma_i\), and non-resonant amplitude \(A_{\mathrm{nr}} = 1\) were fixed during fitting. All other amplitudes and phases were optimized via Bayesian inference.
}
\label{tab:fitparams}
\begin{tabular}{l|c|c|c}
\hline
\textbf{Parameter} & \textbf{Value} & \textbf{Unit} & \textbf{Notes} \\
\hline
$A_1$ & 0.4900 & a.u. & Amplitude \\
$A_2$ & 0.3500 & a.u. & \\
$A_3$ & 0.3265 & a.u. & \\
$A_4$ & 0.8250 & a.u. & \\
$A_5$ & 0.6966 & a.u. & \\
$A_6$ & 1.3300 & a.u. & \\
$A_7$ & 1.5089 & a.u. & \\
$A_8$ & 1.1923 & a.u. & \\
\hline
$\phi_1$ & $0.083~\pi$ & rad & Relative phase \\
$\phi_2$ & $-0.955 \pi$ & rad & \\
$\phi_3$ & $-0.207~\pi$ & rad & \\
$\phi_4$ & $-0.056~\pi$ & rad & \\
$\phi_5$ & $0.923~\pi$ & rad & \\
$\phi_6$ & $0.118~\pi$ & rad & \\
$\phi_7$ & $-0.901~\pi$ & rad & \\
$\phi_8$ & $0.013~\pi$ & rad & \\
\hline
$A_{\mathrm{nr}}$ (fixed) & 1.000 & a.u. & Non-resonant amplitude  \\
$\phi_{\mathrm{nr}}$ & $-0.715~\pi$ & rad & Non-resonant phase \\
\hline

$w_i$ (fixed) & [995, 1003.16, 1038.61, 1080.7, 1278, 1472, 1584, 1596.6] & cm$^{-1}$ & Resonance positions \\
$\gamma_i$ (fixed) & [5, 4, 4, 5, 5, 8, 8, 6] & cm$^{-1}$ & Damping factors \\
\hline
\end{tabular}
\end{table}

\begin{enumerate}
    \item \textbf{Baseline Removal and Normalization::} \\
    Prior to fitting, experimental spectra were baseline-corrected using low-order polynomial subtraction to isolate the resonant components. The baseline was fitted using a fixed-order polynomial (typically 2nd or 3rd order), optimized separately for each dataset to minimize residual curvature.Following baseline correction, all spectra were normalized to the [0,1] range to account for the intrinsic intensity mismatch between SFG and DFG signals, ensuring numerical stability during fitting.

\item \textbf{Fitting Model and Procedure:} \\

The model is based on a sum of complex Lorentzian resonances, representing both resonant contributions, and constant complex number representing the non-resonance (frequency-independent) nonlinearity:

\begin{equation}
X_{\text{SFG}}(x) = \sum_{i=1}^{8} A_i \cdot \frac{e^{j\phi_i}}{w_i - x - j\gamma_i / 2} + e^{j\phi_{\text{nr}}}
\end{equation}

\begin{equation}
X_{\text{DFG}}(x) = \sum_{i=1}^{8} A_i \cdot \frac{e^{j\phi_i}}{w_i - x + j\gamma_i / 2} + e^{j\phi_{\text{nr}}}
\end{equation}

The observable signals are obtained by computing the squared magnitudes:
\[
I_{\text{SFG}} = |X_{\text{SFG}}(x)|^2, \quad I_{\text{DFG}} = |X_{\text{DFG}}(x)|^2, \quad I_{\text{Ratio}} = \frac{I_{\text{SFG}}}{I_{\text{DFG}}}
\]

\paragraph{Single-mode limit of $R(\omega)$.}
Let $\Delta\equiv\omega_{1}-\omega$ and $D(\omega)\equiv\Delta^{2}+(\Gamma/2)^{2}$. Write the effective second-order responses as
\[
\chi^{(2)}_{\mathrm{SFG}}(\omega)=A_{\mathrm{nr,SFG}}+A_{1}\,e^{i\Delta\phi_{\mathrm{SFG}}}\,\frac{1}{\Delta - i\Gamma/2},\quad
\chi^{(2)}_{\mathrm{DFG}}(\omega)=A_{\mathrm{nr,DFG}}+A_{1}\,e^{i\Delta\phi_{\mathrm{DFG}}}\,\frac{1}{\Delta + i\Gamma/2},
\]
with $A_{\mathrm{nr,\*}}\in\mathbb{C}$ and $\Delta\phi_{\*}\equiv\phi_{1}-\phi_{\mathrm{nr,\*}}$. Neglecting slow prefactors, the intensities scale as $I_{\mathrm{SFG}}\propto|\chi^{(2)}_{\mathrm{SFG}}|^{2}$ and $I_{\mathrm{DFG}}\propto|\chi^{(2)}_{\mathrm{DFG}}|^{2}$, yielding
\[
R(\omega)\equiv\frac{I_{\mathrm{SFG}}}{I_{\mathrm{DFG}}}
=\frac{|A_{\mathrm{nr,SFG}}|^{2}+\frac{|A_{1}|^{2}}{D}
+\frac{2|A_{\mathrm{nr,SFG}}||A_{1}|}{D}\left[\Delta\cos\Delta\phi_{\mathrm{SFG}}-\frac{\Gamma}{2}\sin\Delta\phi_{\mathrm{SFG}}\right]}
{|A_{\mathrm{nr,DFG}}|^{2}+\frac{|A_{1}|^{2}}{D}
+\frac{2|A_{\mathrm{nr,DFG}}||A_{1}|}{D}\left[\Delta\cos\Delta\phi_{\mathrm{DFG}}+\frac{\Gamma}{2}\sin\Delta\phi_{\mathrm{DFG}}\right]}.
\]
Far from resonance ($D\to\infty$), $R(\omega)\to |A_{\mathrm{nr,SFG}}/A_{\mathrm{nr,DFG}}|^{2}$. If $A_{\mathrm{nr,SFG}}=A_{\mathrm{nr,DFG}}$ and $\Delta\phi_{\mathrm{SFG}}=-\Delta\phi_{\mathrm{DFG}}$, the leading resonant correction to $R(\omega)$ is antisymmetric in $\Delta$, producing a dispersive zero crossing at $\omega=\omega_{1}$. 

Fitting was performed using Bayesian optimization (implemented via the \texttt{bayes\_opt} package), optimizing over 17 parameters: eight resonance amplitudes ($A_i$), eight corresponding phases ($\phi_i$), and one global non-resonant phase ($\phi_{\mathrm{nr}}$). The resonance frequencies $w_i$ and damping factors $\gamma_i$ were held fixed during fitting. These fixed parameters were determined from experimental Raman spectra, where $w_i$ corresponds to the observed vibrational peak positions, and $\gamma_i$ was estimated from the full width at half maximum (FWHM) of each Raman peak.

\item \textbf{Optimization and Result Interpretation:} \\

The cost function minimized the average squared error between the modeled and measured normalized SFG, DFG, and ratio spectra. The solution is not unique; however, the fitting converges to parameter ranges consistent with physical expectations. The fitted parameters were exported for interpretation, and final curves were compared visually with experimental data (see Fig.~2(c)).

\vspace{1em}
Model implementation was performed in Python using CuPy and SciPy for GPU-accelerated complex arithmetic and fitting. Code is available upon reasonable request.
\end{enumerate}
\vspace{1em}
Analysis scripts are available upon reasonable request.

%\subsection*{Additional Spectra and Control Data}

%%%%%%%%%%%%%%%%%%%%%%%%%%%%%%%%%%%%%%%%%%%%%%%%%%%%%%%%%%%%%%%%%%%%
\section{DFT computational method}

According to the literature, modelling a single thiol molecule by replacing the thiol hydrogen with a gold atom in the simulation improves the DFT description of the experimental SERS spectra of thiol SAMs, leading to better agreement between theory and experiment\cite{humbert2011}.
All quantum chemistry calculations were performed with Gaussian16. In our simulations, we considered both the isolated BPhT molecule and the gold-linked version (Au–BPhT), optimizing their geometries at the def2-TZVPP/B3LYP level with the molecular axis aligned along z. Geometry optimizations employed tight convergence thresholds ($10^{-5}$ Hartree/Bohr for forces and 4 × $10^{-5}$ Bohr for RMS displacements, with maxima 1.5× larger). Ground-state minima were verified via Hessian analysis.

To evaluate IR, SERS, and resonant SFG responses, we computed dipole moment and polarizability derivatives along the normal modes using a three-point finite difference scheme. For each mode, the minimum-energy geometry was distorted by ±0.01 Å; reducing the step to ±0.001 Å yielded equivalent results. Electric dipole vectors and polarizability tensors were calculated for each displaced structure. Frequency-dependent polarizabilities were obtained through dynamic CPKS calculations at an excitation wavelength of 785 nm. For these calculations, the aug-cc-pVTZ basis set was used for H, C, and S atoms, and aug-cc-pVTZ-PP for Au, to account for the higher number of polarization and diffusion functions required for such  kind of calculations.
We restricted the analysis to the z-component of the dipole moment and the ZZ component of the Raman tensor, since in the nanogap the local field is predominantly normal to the gold surface, where BPhT molecules self-assemble nearly perpendicular to form the monolayer. STM measurements from our group confirmed that BPhT adopts an almost upright orientation in Au–SAMs, and testing different orientations did not improve agreement with experiment.
The resonant $\chi^{(2)}$ contribution from BPhT was evaluated following the formalism of Lin \cite{Lin1996}. IR and SERS intensities were calculated as the squared modulus of the dipole derivative (z-component) and the Raman polarizability derivative (ZZ component), respectively, with both corrected by the mode’s Fermi population factor.

DFT also predicts two strong SFG-active modes at 1038~cm$^{-1}$ and 1'080~cm$^{-1}$, which appear much weaker experimentally. Adjusting the orientation of the molecule with respect to the gold plane in the simulations did not reconcile this discrepancy, highlighting the limits of molecular calculations in complex plasmonic enviroments. 
One possible explanation is a modification of the vibrational lifetime due to the enhanced MIR plasmonic near field, as previously proposed in scanning-tip experiments \cite{metzger2019}. The Purcell effect, if strong enough, can indeed cause a lifetime broadening of the IR-active transition, following the same physics as optically-allowed electronic transition of molecules at VIS frequencies \cite{verlekar2025}. However, it remains unclear why other vibrational modes with strong IR dipoles, such as the 1340~cm$^{-1}$ mode of 4-NTP discussed below, are not similarly affected. These discrepancies underscore the difficulties of state-of-the-art DFT models to capture complex metal–molecule and plasmon-vibration interactions. Nevertheless, our experimental SFG/DFG spectra are robust across molecules and metasurfaces, preserving line shapes while suppressing MIR baseline drifts. Combined with the compact cw implementation, this capability positions the platform as a practical bridge between chip-integrated photonics and chemically specific MIR spectroscopy that is sensitive to local chemical and photonic environments.

\end{document}